\documentclass[floats,preprint,aps,tightenlines,
nofootinbib,superscriptaddress]{revtex4}
\usepackage{graphicx}
\usepackage{bm}
\usepackage{epsfig}

\newcommand{\bea}{\begin{eqnarray}}
\newcommand{\eea}{\end{eqnarray}}
\newcommand{\beq}{\begin{equation}}
\newcommand{\eeq}{\end{equation}}
\newcommand{\bay}{\begin{array}}
\newcommand{\eay}{\end{array}}

\newcommand{\nn}{\nonumber}

\begin{document}
\preprint{\vbox{ \hbox{UCSD/PTH 02--17} 
\hbox{INT-PUB 02-43}
\hbox{hep-ph/0208034} 
 }}

\title{Testing Factorization in $B\to D^{(*)}X$ Decays\\ \vspace{0.2cm}}

\author{C.W.~Bauer}
\affiliation{Department of Physics, University of California at San Diego,
La Jolla, CA 92093\footnote{Electronic address: bauer@physics.ucsd.edu,
bgrinstein@ucsd.edu, pirjol@bose.ucsd.edu}\vspace{0.2cm}}

\author{B. Grinstein}
\affiliation{Department of Physics, University of California at San Diego,
La Jolla, CA 92093\footnote{Electronic address: bauer@physics.ucsd.edu,
bgrinstein@ucsd.edu, pirjol@bose.ucsd.edu}\vspace{0.2cm}}

\author{D. Pirjol}
\affiliation{Department of Physics, University of California at San Diego,
La Jolla, CA 92093\footnote{Electronic address: bauer@physics.ucsd.edu,
bgrinstein@ucsd.edu, pirjol@bose.ucsd.edu}\vspace{0.2cm}}

\author{I.W. Stewart\vspace{0.3cm}}
\affiliation{Institute for Nuclear Theory,  University of Washington, Seattle, 
	WA 98195\footnote{Electronic address: iain@phys.washington.edu}
	\vspace{0.5cm}}

\vspace{1.0cm}

\begin{abstract}
\vspace{0.3cm} 

In QCD the amplitude for $\bar B^0\to D^{(*)+}\pi^-$ factorizes in the
large $N_c$ limit or in the large energy limit $Q\gg\Lambda_{\rm QCD}$
where $Q=\{m_b,m_c,m_b-m_c\}$. Data also suggests factorization in the
processes $B \to D^* \pi^+ \pi^- \pi^- \pi^0$ and $B \to D^*
\omega \pi^-$, however by themselves neither large $N_c$ nor large $Q$
can account for this. 
Noting that the condition for large energy
release in $\bar B^0\to D^+\pi^-$ is enforced by the SV limit, $m_b\gg
m_b-m_c\gg\Lambda$, we propose that the combined large $N_c$ and SV
limits justify factorization in $B \to D^{(*)} X$. This combined limit is
tested with the $B\to D^*X$ inclusive decay spectrum measured by CLEO.
We also give exact large $N_c$ relations among isospin 
amplitudes for $\bar B\to D^{(*)}X$ and $\bar B\to D^{(*)}\bar D^{(*)}X$, 
which can be used to test factorization through exclusive or inclusive
measurements. 
Predictions for the modes $B\to D^{(*)}\pi\pi$, $B\to D^{(*)}K\bar K$ and
$B\to D^{(*)}\bar D^{(*)} K$ are discussed using available data.

\end{abstract} 
\maketitle

\section{Introduction}

There are two limits of QCD in which factorization in $\bar B^0\to
D^+\pi^-$ (and related) decays can be proven rigorously. In the large $N_c$ limit,
in which one takes the limit of infinite number of colors, one can
show that all nonfactorizable contributions are suppressed by
$1/N_c^2$. In the large energy limit, $Q \gg \Lambda_{\rm QCD}$,
where $Q$ denotes any one of the large scales $m_b$, $m_c$ or $E_\pi$,
factorization has been proven rigorously~\cite{bpsfact} using a
soft-collinear effective theory~\cite{scet}. In this case the
corrections to factorization are suppressed by powers of
$1/Q$~\cite{Dugan:1990de}.

The large $Q$ proof of factorization has its roots in a non-rigorous
argument by Bjorken that justifies factorization using color
transparency. Here the requirement is that the pion velocity in the
$B$ rest frame be ultrarelativistic, so that the small color dipole of
quarks that forms the $\pi^-$ grows to hadronic size $\Lambda$ only
after a time delayed hadronization time $\gamma\Lambda$ much larger
than the hadronic size of the $B$ meson. Although not systematic,
color transparency therefore seems to suggest that violations to
factorization are order $1/\gamma$.  
While both large $N_c$ and large $Q$ lead to factorization in a
particular limit of QCD, it is not clear which is the most important
for the phenomenologically observed factorization in real QCD, in
which neither large $N_c$, large $Q$, nor large $\gamma$ are exactly
realized. Each method gives similar predictions for the decay $B^0 \to
D^+ \pi^-$ and, while none predict the decay $B^0 \to \bar D^0
\pi^0$, they all render it suppressed. Hence these decays are not very
useful for distinguishing among these arguments for factorization.

If large $N_c$ were the only requirement for factorization, one would
expect factorization to hold equally well in $D$ decays and $K$
decays. Factorization does not work very well in $D$ decays \cite{BSW,NeSt}
and is even more strongly
violated in $K$ decays. Clearly large $N_c$ by itself does not explain
phenomenological factorization. A great deal of effort has been devoted
to explaining this aparent puzzle \cite{factor1,factor2,factor3}.

On the other hand, if the large energy limit was solely responsible 
for explaining
factorization in $B$ decays, one would expect \cite{Dugan:1990de}
corrections to grow as $m_X/Q$, where $m_X$ is the hadronic mass
produced by the $(\bar d u)_{V-A}$ current. Recently, an analysis of
the decays $B \to D^* \pi^+ \pi^- \pi^- \pi^0$ and $B \to D^* \omega
\pi^-$ was performed, investigating the applicability of factorization
as a function of the invariant mass of the light hadrons
\cite{Ligeti:2001dk}. Using information from $\tau$ decays
it was shown that up to $m_X \sim 1.7$ GeV there is no indication of
violations to factorization, which indicates that the large $Q$ limit
can not be solely responsible for explaining factorization.  However,
this data can not be explained solely from the large $N_c$ limit of
QCD either. In order to calculate the predictions from factorization,
Ref.~\cite{Ligeti:2001dk} used the factorized form
\begin{eqnarray} \label{fact1}
  \langle X D^{(*)} | (\bar c b)_{V-A}(\bar d u)_{V-A} | B \rangle 
  =\langle D^{(*)} | (\bar c b)_{V-A} | B \rangle 
  \langle X |(\bar d u)_{V-A} | 0 \rangle\,.
\end{eqnarray}
This neglects the possibility that part of the hadronic state
containing the light particles can be created from the $b \to c$
current. In fact, large $N_c$ only gives
\begin{eqnarray} \label{fact2}
\langle X D^{(*)} | (\bar c b)_{V-A}(\bar d u)_{V-A} | \bar B \rangle 
= 
\sum_{X',X''} \langle D^{(*)}X' | (\bar c b)_{V-A} | B \rangle 
\langle X''| (\bar d u)_{V-A} | 0 \rangle\,,
\end{eqnarray}
with $X'$ and $X''$ adding to give the final light hadron state
$X$. The authors of Ref.~\cite{Ligeti:2001dk} addressed this issue by
arguing that the contributions from non-zero $X'$ may be numerically
small and that for $B\to D^*\omega \pi^-$ this can be tested using
data from semileptonic $B \to D \omega \ell \bar \nu$ decays.
These effects can also be tested in the decay
$\bar B^0\to D^{*0} \pi^+ \pi^+ \pi^- \pi^-$ for which $X'=\{\pi^+\,,\pi^+\pi^+\pi^-
\}$ and
$X'' = \{\pi^+\pi^-\pi^-,\pi^-\}$. The CLEO measurement \cite{CLEO4pi} finds
a small but nonvanishing branching ratio for this mode 
$Br(\bar B^0\to D^{*0} \pi^+ \pi^+ \pi^- \pi^-) = (0.30 \pm 0.07 \pm 0.06)\%$.

It is interesting to note that factorization in
Eq.~(\ref{fact1}) is justified in the small velocity (SV) limit,
$m_b,m_c\gg m_b-m_c\gg\Lambda_{\rm QCD}$ \cite{SV}. As shown in
Ref.~\cite{Boyd:1995ht}, in the SV limit the inclusive $B \to X_c \ell
\bar \nu$ branching ratio is saturated by $D$ and $D^*$ final
states. Therefore, in the SV limit
\begin{eqnarray}\label{SVnoX}
 \langle X' D^{(*)} | (\bar c b)_{V-A} | B \rangle  = \langle
 D^{(*)} | (\bar c b)_{V-A} | B \rangle \delta_{X',0}\,.
\end{eqnarray}
Thus, Eq.~(\ref{fact2}) reduces to Eq.~(\ref{fact1}) in the simultaneous limit 
of large $N_c$ and SV and this combined limit is therefore capable of 
explaining the results of \cite{Ligeti:2001dk}. 

Although more stringent
than the large $Q$ limit, the $SV$ limit implies large $Q$ and is
therefore fully consistent with it and also gives additional
predictive power. For $B\to D^{(*)}$ transitions the relevant
kinematic limits can be summarized as
\begin{eqnarray}\label{limits}
  \mbox{HQET: } && m_b,m_c \gg \Lambda_{\rm QCD} \,,\nn\\
  \mbox{Large $Q$: } && m_b,m_c,m_b-m_c \gg \Lambda_{\rm QCD} \,,\nn\\
  \mbox{$SV$ limit: } && m_b,m_c \gg m_b-m_c \gg \Lambda_{\rm QCD} \,,
\end{eqnarray}
each of which is a subset of the one above. The requirement that the
light degrees of freedom in the $D^{(*)}$ can be described by HQET
requires that $E_D\sim m_c$ or equivalently that the $B$ and $D$
velocities have $v\cdot v'$ of order one. This implies that $m_b-m_c
\sim \sqrt{m_b m_c}$ or $m_b-m_c \ll \sqrt{m_b m_c}$, which are
allowed scalings in the large $Q$ and SV limits respectively.

The purpose of this paper is twofold. In section \ref{isospin} we show 
that even the more generally factorized form of the amplitude in 
Eq.~(\ref{fact2}) leads to experimentally testable predictions which
are distinct from those following from large $Q$ factorization.
The $B\to D^{(*)}X_u$ decays with charge eigenstates are parameterized by 
four independent isospin amplitudes.
In the large $N_c$ limit, they are given in terms of only
two reduced matrix elements, compared with just one in large $Q$ factorization. 
A similar result holds for $B\to D\bar D X$
decays, for which we prove a similar reduction in isospin amplitudes from 7 to 5,
in contrast with two in the combined large $N_c$ and SV  factorization.
In section \ref{simlim} we consider the 
simultaneous limit of large $N_c$ and $SV$ and calculate
the inclusive differential decay rate $B \to D^* X$ in
this limit. The results obtained are compared with data available
from the CLEO collaboration~\cite{Gibbons:1997ag}.

\section{Large $N_c$ relations for $B\to DX$}
\label{isospin}
The final state $X$ in the decay $\bar B \to D X$ may have charm number
$-1$ or 0, depending on whether the underlying weak decay is $b\to c
\bar c s $ or $b\to c \bar u d $. We analyze these separately,
correspondingly labeling the final states $X_c$ and $X_u$.  
Taking into account the fact that the $B$ and
$D$ states belong to isospin doublets and that the weak Hamiltonian
\begin{eqnarray} \label{Hw}
 {\cal H}_W &=& \frac{G_F}{\sqrt{2}} V_{cb} V_{ud}^* \left[C_1(\mu)\,
  (\bar c u)_{V-A} (\bar d b)_{V-A} + C_2(\mu)\, (\bar c b)_{V-A} 
  (\bar d u)_{V-A} \right]
\end{eqnarray}
transforms as an isotriplet with 
$(I,I_3)=(1,+1)$, one obtains 
that the final light state $X_u$ can have isospin $I_{X_u} = 0, 1, 2$. 
Specifying the isospin of the $X_u$ system together with the total isospin 
of the final hadronic system gives rise to four isospin states: 
$[DX_0]_{\frac12}$, $[DX_1]_{\frac12}$, $[DX_1]_{\frac32}$ and $[DX_2]_{\frac32}$. 
The corresponding reduced isospin amplitudes are
\bea\label{isoamps}
& &a_{\frac12} = \langle [DX_0]_{\frac12}|\!|{\cal H}_W|\!|\bar B \rangle\,,\quad
b_{\frac12,\frac32} 
= \langle [DX_1]_{\frac12,\frac32}|\!|{\cal H}_W|\!|\bar B\rangle\,,\quad
c_{\frac32} = \langle [DX_2]_{\frac32}|\!|{\cal H}_W|\!|\bar B\rangle\,.
\eea
The isospin amplitudes 
depend on $X$ through its hadronic content and particle momenta.
Squaring the amplitudes and summing over the isospin of the $X_u$ state one 
finds for the rates with $B$ and $D$ charge eigenstates
\bea
& &\Gamma(\bar B_d\to D^+ X_u^-) = \sum_X \left| \frac23 b_{1/2}(X) + 
\frac13 b_{3/2}(X) \right|^2 +  \sum_X \frac15 |c_{3/2}(X)|^2\nonumber\\
& &\Gamma(\bar B_d\to D^0 X_u^0) = \sum_X \frac23 |a_{1/2}(X)|^2 + 
\sum_X \frac29 \left| - b_{1/2}(X) + 
 b_{3/2}(X) \right|^2 +  \sum_X \frac{2}{15} |c_{3/2}(X)|^2\nonumber\\
& &\Gamma(B^-\to D^0 X_u^-) = \sum_X \left| b_{3/2}(X) \right|^2 
+  \sum_X \frac15 |c_{3/2}(X)|^2\nonumber\\
& &\Gamma(B^-\to D^+ X_u^{--}) = \sum_X \frac45 |c_{3/2}(X)|^2\,.
\label{rt1}
\eea
The phase space factors are implied. 
The corresponding sums over $X$ in the rate formulas
include summation over the final states and integration over the
phase space.

In the large $N_c$ limit these amplitudes simplify considerably. 
We assign the usual $N_c$ power counting to the Wilson coefficients
$C_1(m_b)\sim 1/N_c$, $C_2(m_b)\sim 1$, in agreement with their
perturbative expansion in $\alpha_s$ \cite{NeSt}.
This gives that to leading order in $1/N_c$ the $\bar B\to DX$ amplitude
factors as
\bea\label{A_Nc}
A_{\rm N_c^0}(\bar B\to DX_u ) = \frac{G_F}{\sqrt2} V_{cb} V_{ud}^* C_2
\sum_{X',X''}\langle DX' |\bar c b|\bar B\rangle \langle X'' |\bar du| 0\rangle +
O(1/N_c) \eea
with $X_u=X'+X''$. 
Since the state $X''$ has to be in a state of isospin $|I,I_3\rangle = |1,
-1\rangle$ and the state $X'$ can only have isospin $I= 0$ or 1, all 
possible decays $\bar B\to DX_u$ are
determined in the large $N_c$ limit by two amplitudes $F_I$ defined as
\bea
F_0(X) &=& \sum_{X' X''} 
\langle [DX'_0]_{\frac12}|\!|\bar c b|\!|\bar B\rangle \langle X''_1|\!|\bar d u|\!|0\rangle\,,\\
F_1(X) &=& \sum_{X' X''} 
\langle [DX'_1]_{\frac12}|\!|\bar c b|\!|\bar B\rangle \langle X''_1|\!|\bar d u|\!|0\rangle\,.
\nonumber
\eea
The four isospin amplitudes (\ref{isoamps})
can be written now in terms of $F_0$ and $F_1$ as\footnote{Although in general
$F_{0,1}(X)\sim O(N_c^0)$, for special states $X$ some of these amplitudes may vanish.
For example, taking $X=\pi$ gives $F_1(\pi) = 0$.}
\bea\label{largeNisospin}
a_{1/2} = -F_1 \,,\qquad b_{1/2} = F_0 + \sqrt2 F_1 \,,\qquad
b_{3/2} = F_0 - \frac{1}{\sqrt2} F_1\,,\qquad c_{3/2} = \sqrt{\frac52}
F_1\,.  
\eea

Using these expressions in the rate formulas, one finds the
large $N_c$ predictions
\bea & &\Gamma_{\rm N_c^0}(\bar B_d\to D^+ X_u^-) = \sum_X |F_0(X)
+ \frac{1}{\sqrt2} F_1(X)|^2 + \frac12|F_1(X)|^2\nonumber\\ 
& &\Gamma_{\rm N_c^0}(\bar
B_d\to D^0 X_u^0) = \sum_X 2|F_1(X)|^2\nonumber\\
& &\Gamma_{\rm N_c^0}(B^-\to D^0 X_u^-) = \sum_X |F_0(X) - \frac{1}{\sqrt2} F_1(X)|^2 
+ \frac12|F_1(X)|^2\nonumber\\
\label{largeN}
& &\Gamma_{\rm N_c^0}(B^-\to D^+ X_u^{--}) = \sum_X 2|F_1(X)|^2\,.
\eea
The two independent amplitudes can be extracted from the combinations of rates
\bea\label{18a}
& &\Gamma(\bar B_d \to D^+ X_u^-) + \Gamma(B^-\to D^0 X_u^-) = \sum_X 2|F_1(X)|^2 
+ 2|F_0(X)|^2\\
\label{18}
& &\Gamma(\bar B_d \to D^0 X_u^0) = \Gamma(B^-\to D^+ X_u^{--}) = \sum_X 2|F_1(X)|^2\,.
\eea

The corrections to the predictions in Eq.~(\ref{largeNisospin})
come at order $1/N_c$ and are
parameterized by the subleading term in Eq.~(\ref{A_Nc}). This can be
written again in a factorized form as
\bea\label{O(1/Nc)}
A_{1/N_c}(\bar B\to DX_u ) = \frac{G_F}{\sqrt2} V_{cb} V_{ud}^* 
\left(C_1+\frac{C_2}{N_c}\right)
\sum_{X',X''}\langle DX' |\bar c u|0\rangle \langle X'' |\bar db|\bar B\rangle +
O(1/N_c^2)\,. 
\eea
For this case both $X'$ and $X''$ can have isospin $I=0,1$. 
Thus four isospin amplitudes are required to describe all matrix elements 
(\ref{O(1/Nc)}), which shows that in general no reduction in the number of isospin 
amplitudes persists beyond leading order in $1/N_c$.

The large energy limit can be applied to these decays if the final state $X$ contains 
only particles 
which form a jet with energy much larger than its invariant mass (this includes the case of 
just a single light particle with large energy). In this case, the amplitude can be factored 
in a way similar to the large $N_c$ limit as
\bea\label{A_E}
A_{\rm E}(\bar B\to DX_u ) = \frac{G_F}{\sqrt2} V_{cb} V_{ud}^* \left(C_2+\frac{C_1}{N_c}\right)
\langle D |\bar c b|\bar B\rangle \langle X |\bar du| 0\rangle +
{\cal O}\left(\alpha_s(Q),\left(\frac{m_X}{Q}\right)^n\right) 
\eea
where $m_X$ is the invariant mass of the state $X$ and $n>0$.
Since the isospin of $X$ is constrained to be $I=1$, the only nonvanishing isospin
amplitudes in this limit are $b_{1/2}, b_{3/2}$.
This prediction can be tested for example by measuring the decay 
$B^- \to D^+ \pi^- \pi^-$ as a function of the angle between the two energetic pions, 
which should vanish as this angle decreases.

The above analysis can be carried through in an identical manner
for $B\to D^*X$ decays.
In the combined large $N_c$ and SV limit the amplitude $\langle D^{(*)}X |\bar c b|\bar B\rangle$ 
vanishes for $X\neq 0$. 
Thus, $F_1(X) = 0$ for all final states $X$. This prediction is similar to the one from large 
energy factorization, but without the kinematic requirement on the light hadronic system $X$. 
Separate measurements of the 4 rates with $B,D$ charge eigenstates would allow
a test of these predictions, and distinguish between large $N_c$ factorization, large energy 
factorization or the combined limit of large $N_c$ and $SV$.
In the next section we discuss a partial test along these lines,
making use of the present limited experimental information on $B\to D^{(*)} X$
available from CLEO \cite{Gibbons:1997ag}.

We stress that the large $N_c$ relations hold not only 
for the inclusive mode, but also for states $X$ with fixed hadronic content. 
For example, taking $X=\pi\pi$ gives that in the large $N_c$ limit, the 
amplitudes for the two pions in $\bar B\to D^{(*)}(\pi\pi)_I$ to be emitted in states of 
isospin $I=0$ and 2 are related. Eq.~(\ref{18}) gives a relation among rates 
to leading order in $1/N_c$
(note that $\Gamma(\bar B_d\to D^{(*)0}\pi^0\pi^0) \sim O(1/N^2_c)$)
\bea\label{largeN2pi}
\Gamma(\bar B_d\to D^{(*)0}\pi^+\pi^-) =
\Gamma( B^-\to D^{(*)+}\pi^-\pi^-) + O(1/N_c)\,.
\eea
The branching ratios of these modes have been recently reported by the 
BELLE Collaboration \cite{BelleICHEP}. For the $D$ modes they are 
$Br(\bar B_d\to D^0\pi^+\pi^-) = (7.5 \pm 0.7 \pm 1.5)\times 10^{-4}$ and
$Br(B^-\to D^+\pi^-\pi^-) = (1.07 \pm 0.04 \pm 0.16)\times 10^{-3}$,
and for the $D^*$ modes
$Br(\bar B_d\to D^{*0}\pi^+\pi^-) = (6.2 \pm 1.2 \pm 1.7)\times 10^{-4}$,
$Br(B^-\to D^{*+}\pi^-\pi^-) = (1.24 \pm 0.07 \pm 0.22)\times 10^{-3}$.

The data agrees fairly well with the large $N_c$ prediction for the
$D$ modes and suggests larger $1/N_c$ corrections for the $D^*$
modes. This agreement permits a determination of $|F_1(\pi\pi)|^2$
(integrated over phase space) using Eq.~(\ref{18}). A subtraction can
then be performed to extract $|F_0(\pi\pi)|^2$ using (\ref{18a}). The
widths on the left side of Eq.~(\ref{18a}) are dominated by the
two-body $B\to D\rho$ mode\cite{Alam:bi}.  One finds (in units of
$Br$)
\bea
 2|F_0(\pi\pi)|^2 = (20.4\pm 2.3)\times 10^{-3}\,,\qquad
 2|F_1(\pi\pi)|^2 = (9.1 \pm 1.1)\times 10^{-4}\,.
\eea
The large observed enhancement of $F_0$ over $F_1$ is a consequence of
the inequality among two-body modes $\Gamma(D\rho) \gg\Gamma(D^{**}\pi)$. 
Factorization for these two body modes can be explained either in large 
$N_c$ or in the large $Q$ limit. The suppression of $F_1$ over $F_0$  
is in agreement with the SV limit.

Another test of the large $N_c$ relations is obtained by taking $X=K\bar K$,
for which data is available from BELLE \cite{BELLE2K}. The isospin of the kaon
pair can be only 0 and 1, which gives $c_{3/2}(K\bar K)=0$. The large $N_c$ relations 
(\ref{largeNisospin}) imply $F_1(K\bar K) = 0$, which shows that in this limit the
$B\to D(K\bar K)_{I=0}$ amplitude is suppressed, and the decay rates satisfy
\bea
& &\Gamma(\bar B_d \to D^+ K^-\bar K^0) = \Gamma(B^-\to D^0 K^-\bar K^0 ) + O(1/N_c)\nonumber\\
& &\Gamma(\bar B_d \to D^0 K^0\bar K^0) = \Gamma(\bar B_d\to D^0 K^+ K^- ) = O(1/N_c^2)\,.
\eea
The first prediction agrees well with the BELLE results \cite{BELLE2K}
$Br(\bar B_d \to D^+ K^-\bar K^{*0})
= (8.8 \pm 1.1 \pm 1.5)\times 10^{-4}$, $Br(B^- \to D^0 K^-\bar K^{*0}) = (7.5\pm
1.3 \pm 1.1)\times 10^{-4}$ and 
$Br(\bar B_d \to D^{*+} K^-\bar K^{*0})
= (12.9 \pm 2.2 \pm 2.5)\times 10^{-4}$, $Br(B^- \to D^{*0} K^-\bar K^{*0}) = (15.3\pm
3.1 \pm 2.9)\times 10^{-4}$.

Next we turn our attention to the $b\to c\bar c s$ process. 
For this case, large $Q$ arguments can not be used to justify 
factorization, which leaves large $N_c$ as the sole possible
explanation. The Hamiltonian ${\cal H}^{cs}_W$ responsible for these
decays is identical to (\ref{Hw}) with the substitution $u\to c$. We neglect the
penguin operators because of their small CKM factors and Wilson coefficients.
We first
consider the case in which experiments may tag the $B$ meson and
distinguish the charm from anti-charm. 
There are two isospin amplitudes
$h_0, h_1$  corresponding to the two possible values of the isospin $I$ 
of the state $X_c$
\bea
h_0(X) = \langle [DX_{c0}]_{\frac12}|\!|{\cal H}^{cs}_W|\!|\bar B \rangle\,,\qquad
h_1(X) = \langle [DX_{c1}]_{\frac12}|\!|{\cal H}^{cs}_W|\!|\bar B \rangle\,.
\eea
The corresponding rates are given by
\bea
& &\Gamma(\bar B_d\to D^+ X_c^-) = \sum_X |h_0(X)|^2 + \frac13|h_1(X)|^2\nonumber\\
& &\Gamma(\bar B_d\to D^0 X_c^0) = \sum_X \frac23|h_1(X)|^2\nonumber\\
& &\Gamma(B^-\to D^+ X_c^{--}) = \sum_X  \frac23|h_1(X)|^2\nonumber\\
\label{bccs}
& &\Gamma(B^-\to D^0 X_c^-) = \sum_X |h_0(X)|^2 + \frac13|h_1(X)|^2\,.
\eea
No simplifications are expected
for these modes in the large $N_c$ limit.
Due to the identical isospin structure of their Hamiltonian, similar relations 
can be written down for the semi-inclusive semileptonic decays $\bar B\to DXe\bar\nu$,
in terms of another two amplitudes $g_{0,1}(X)$. 

The definition of the SV limit for decays containing two charm quarks in the final
state is somewhat different from the one introduced in Eq.~(\ref{limits}). Requiring
that both charmed hadrons move slowly gives \cite{factor3}
\bea
  \mbox{Generalized $SV$ limit: } && m_b,m_c \gg m_b-2m_c \gg \Lambda_{\rm QCD} \,.
\eea
In this combined large $N_c$ and
SV limit, the state $X_c$ is produced by the $(\bar sc)$ current and must have isospin
0, which requires the amplitude $h_1(X)$ to vanish $h_1(X) \to 0$.

\begin{table}[t!]
\bea\nonumber
\begin{array}{cc|c|c|c|c|c}
\hline
\hline
\bar B_d \to & D^0\bar D^0 X_{1/2} & D^0\bar D^0 X_{3/2} &
               D^0 D^- X_{1/2}     & D^0 D^- X_{3/2} & 
               D^0 D_s^- X_0    & D^0 D_s^- X_1\\
             & \sqrt{\frac23} a_2 & -\frac{1}{\sqrt6}a_3 &
               0 & \frac{1}{\sqrt2}a_3 & 
               0 & \sqrt{\frac23} b_2 \\
\hline
             & D^+ \bar D^0 X_{1/2} & D^+ \bar D^0 X_{3/2} &
               D^+ D^- X_{1/2} & D^+ D^- X_{3/2} & 
               D^+ D_s^- X_0 & D^+ D_s^- X_1 \\
             & -\frac{1}{\sqrt6}a_2 + \frac{1}{\sqrt2}a_1 & \frac{1}{\sqrt6}a_3 &
               -\frac{1}{\sqrt6}a_2 - \frac{1}{\sqrt2}a_1 & -\frac{1}{\sqrt6}a_3 &
               b_1 & -\sqrt{\frac13} b_2 \\
\hline
             & D_s^+ \bar D^0 X_0 & D_s^+ \bar D^0 X_1 & 
               D_s^+ D^- X_0 &  D_s^+ D^- X_1 & 
               D_s^+ D_s^- X_{1/2} & \\
             & c_1 & -\sqrt{\frac13}c_2 & 0 & \sqrt{\frac23}c_2 & d &  \\
\hline\hline
B^- \to      & D^0\bar D^0 X_{1/2} & D^0\bar D^0 X_{3/2} & 
               D^0 D^- X_{1/2}     & D^0 D^- X_{3/2} &
               D^0 D_s^- X_0       & D^0 D_s^- X_1 \\
             & \frac{1}{\sqrt6}a_2 + \frac{1}{\sqrt2}a_1 & -\frac{1}{\sqrt6}a_3 &
               \frac{1}{\sqrt6}a_2 - \frac{1}{\sqrt2}a_1 & \frac{1}{\sqrt6}a_3 &
               b_1     &   \frac{1}{\sqrt3} b_2 \\
\hline
             & D^+ \bar D^0 X_{1/2} & D^+ \bar D^0 X_{3/2} &
               D^+ D^- X_{1/2}      & D^+ D^- X_{3/2} & 
               D^+ D_s^- X_0 & D^+ D_s^- X_1 \\
             & 0 & \frac{1}{\sqrt2}a_3 &
               -\sqrt{\frac23} a_2 & -\frac{1}{\sqrt6}a_3 &
               0 & -\sqrt{\frac23} b_2 \\
\hline
             & D_s^+ \bar D^0 X_0 & D_s^+ \bar D^0 X_1 &  
               D_s^+ D^- X_0      & D_s^+ D^- X_1 & 
               D_s^+ D_s^- X_{1/2} & \\
 & 0 & -\sqrt{\frac23} c_2 & c_1 & \sqrt{\frac13}c_2 & d &  \\
\hline\hline
\end{array}
\eea
\begin{quote} {\bf Table I.} Isospin decomposition of the most
general $\bar B\to D\bar DX_I$ decay amplitude into states $X_I$
with isospin $I$.
\end{quote}
\end{table}

Making explicit the charm and anticharm in the final states gives many more
modes. This type of analysis is necessary if the experimental inclusive 
measurement relies on the presence of a charmed meson and separates them 
only according to whether they are charged or neutral. 
Taking into account the fact that the Hamiltonian responsible for these decays
is an isospin singlet $I=0$, 
one finds seven independent isospin amplitudes describing these
decays. Three reduced amplitudes describe decays into nonstrange $D$
mesons
\bea
a_1 = \langle D(\bar DX_{1/2})_0|\!|{\cal H}^{cs}_W|\!|\bar B\rangle\quad
a_2 = \langle D(\bar DX_{1/2})_1|\!|{\cal H}^{cs}_W|\!|\bar B\rangle\quad
a_3 = \langle D(\bar DX_{3/2})_1|\!|{\cal H}^{cs}_W|\!|\bar B\rangle
\eea
and another four reduced amplitudes parameterize decays into $D_s$
\bea
& &b_1 = \langle D(D^-_s X_0)_0|\!|{\cal H}^{cs}_W|\!|\bar B\rangle\qquad
b_2 = \langle D(D^-_s X_1)_1|\!|{\cal H}^{cs}_W|\!|\bar B\rangle\nonumber\\
& &c_1 = \langle \bar D( D^+_s X_0)_0|\!|{\cal H}^{cs}_W|\!|\bar B\rangle\qquad
c_2 = \langle \bar D(D^+_s X_1)_1|\!|{\cal H}^{cs}_W|\!|\bar B\rangle\,.
\eea
Finally, another amplitude $d$ describes $\bar B\to D^+_s D^-_s X_{1/2}$
decays. The isospin decomposition of the most general 
$B \to D \bar D X$ decay amplitude is shown in Table I.

In the large $N_c$ limit, the amplitudes for $B\to D\bar DX$ factor
in a similar way as for $\bar B\to DX_u$ (\ref{A_Nc}), (\ref{O(1/Nc)}).
Keeping terms up to $O(1/N_c)$, the factorizable terms read
\bea
A(\bar B\to D\bar DX) &=& \frac{G_F}{\sqrt2}V_{cb} V_{cs}^*
C_2 \sum_{X',X''}\langle DX'|\bar cb|\bar B\rangle \langle \bar D X''|\bar sc|0\rangle\\
&+&  \frac{G_F}{\sqrt2}V_{cb} V_{cs}^*
\left(C_1 + \frac{C_2}{N_c}\right) 
\sum_{X',X''}\langle D\bar DX'|\bar cc|0\rangle \langle X''|\bar sb|\bar B\rangle 
+ O(1/N_c^2)
\nonumber
\eea
Taking into account isospin constraints on the intermediate states $X',X''$, one finds
that the 3 amplitudes $a_{1-3}$ are given by 2 independent amplitudes $A_{0,1}$ at $O(N_c^0)$
(corresponding to $X'$ in the first term having isospins $I=0,1$) and another 2 amplitudes 
$B_{0,1}$ at $O(1/N_c)$ (corresponding to $X'$ in the second term having isospins $I=0,1$)
\bea\label{a123}
\hspace{-0.5cm}
a_1  = \sqrt2 A_0 - \frac{1}{2N_c}(B_0 + B_1)\,,\quad a_2 = A_1 - \frac{\sqrt3}{2N_c}(B_0-\frac13 B_1)\,,
\quad a_3 = -\sqrt2 A_1 + \frac{1}{N_c}\sqrt{\frac23} B_1\,.
\eea
At leading order in $1/N_c$ there are 2 relations among amplitudes
\bea\label{largeNDDX}
a_3(X) = -\sqrt2 a_2(X)\,,\qquad\qquad  c_1(X) =-\frac{1}{\sqrt3} c_2(X)\,.
\eea
No such relations exist among the coefficients $b_i$.
These predictions imply the large $N_c$ rate relations 
(separately for each of the
four channels $\bar B\to D^{(*)}\bar D^{(*)}X$ and summed over the isospin
of $X$)
\bea
& &\hspace{-0.5cm}\Gamma(\bar B_d\to D^0\bar D^0 X) =
\Gamma(\bar B_d\to D^0 D^- X) =
\Gamma(B^- \to D^+ \bar D^0 X) =
\Gamma(B^- \to D^+ D^- X)\nonumber\\
& &\hspace{-0.5cm}\Gamma(\bar B_d\to D_s^+\bar D^0 X) =
\Gamma(\bar B_d\to D_s^+ D^- X) =
\Gamma(B^- \to D_s^+ \bar D^0 X) =
\Gamma(B^- \to D_s^+ D^- X)\,.\nonumber\\
\eea

\begin{table}[t!]
\bea\nonumber
\begin{array}{c|c|c}
\hline\hline
\mbox{Mode} & \mbox{Decay rate} & \mbox{Br}\times 10^{-3} \\
\hline
\begin{array}{c}
\bar B_d\to D^+\bar D^0 K^- \\
B^- \to D^0 D^- \bar K^0 
\end{array} & |A_0(K)|^2 + O(1/N^2_c) &
\begin{array}{c}
 (1.7 \pm 0.3 \pm 0.3) \\
 (1.8\pm 0.7\pm 0.4) 
\end{array} \\
\hline
\begin{array}{c}
\bar B_d\to D^+ D^- \bar K^0 \\
B^- \to D^0 \bar D^0 K^-
\end{array} & |A_0(K) - \frac{1}{\sqrt2 N_c}B_0(K)|^2 + O(1/N^2_c) &
\begin{array}{c}
(0.8^{+0.6}_{-0.5}\pm 0.3) \\
(1.9 \pm 0.3 \pm 0.3) 
\end{array} \\
\hline
\begin{array}{c}
\bar B_d\to D^0\bar D^0 \bar K^0 \\
B^- \to D^+ D^- K^-
\end{array} & \frac{1}{2N_c^2} |B_0(K)|^2 &
\begin{array}{c}
(0.8\pm 0.4 \pm 0.2) \\
(0.0\pm 0.3 \pm 0.1) 
\end{array} \\
\hline
\hline
\end{array}
\eea
\begin{quote} {\bf Table II.} Large $N_c$ predictions for a few $\bar B\to D\bar DK$
modes. The second column gives the large $N_c$ prediction and the
last column shows the branching fractions for the corresponding modes
measured by the Babar Collaboration \cite{BabarDDK}. The consecutive pairs
of modes have equal widths in the isospin limit.
\end{quote}
\end{table}

No data exists at present
on any of these inclusive modes. It is possible to test the large
$N_c$ predictions on the exclusive modes $\bar B\to D^{(*)} \bar D^{(*)} K^{(*)}$
using data recently available from Babar \cite{BabarDDK}. Taking $X=K^{(*)}$ requires 
$a_3(K)=0$ since the isospin $3/2$ is not allowed. The large $N_c$ relation 
(\ref{largeNDDX}) predicts $a_2(K) = O(1/N_c)$, which implies the rate
relations shown in Table II. 
Comparing these predictions from large $N_c$ with
the data one finds reasonable good agreement within the theoretical and experimental
uncertainties.

The large $N_c$ predictions (\ref{largeNDDX}) can be again contrasted with the more
constraining results obtained in the combined large $N_c$ and SV limits, 
according to which, among the isospin amplitudes $a_i, b_i, c_i, d$, the
only nonvanishing ones are $a_1$ and $b_1$. For the modes in Table II
the SV limit predicts a vanishing $B_0$. However, since this amplitude is
subleading in $1/N_c$, these modes can not be used to test the SV limit.
Therefore separate measurements of other modes in Table I are needed to distinguish 
which of these limits (if any) are actually realized in nature.

\section{Factorization for $\bar B\to D^{*}X$}
\label{simlim}

As a further test of factorization we propose studying the inclusive
hadronic decay $B\to D^*X$ as a function of the invariant mass of the
state $X$, $q^2=m_X^2$.  CLEO has
measured this decay spectrum from $m_X^2\simeq 0$ out to the maximum
hadronic mass $m_X^2=(m_B-m_{D^*})^2\simeq 10.7\,{\rm
GeV}^2$~\cite{Gibbons:1997ag}.  Thus, the inclusive spectrum allows
for a test of factorization over a much larger range than the
$m_X^2\simeq 0$ to $3.2\,{\rm GeV}^2$ region considered in
Ref.~\cite{Ligeti:2001dk}.

As explained in the introduction, the inclusive $B\to D^*X$ amplitude in the large 
$N_c$ limit is given by
\begin{equation} \label{ADX}
 {\cal A}(\bar B\to D^*X)= \frac{G_F}{\sqrt2}V_{ud}^*V_{cb}\sum_{X',X''} C_2(\mu) 
 \langle D^{*}X' | (\bar c b)_{V-A} | B \rangle 
 \langle X''| (\bar d u)_{V-A} | 0 \rangle\,,
\end{equation}
with a sum over $X=X'+X''$ in the rate, plus additional analogous
contributions involving the $(\bar s c)_{V-A}$ current and Cabibbo
suppressed terms.  Imposing the SV limit as well gives rise to two important 
simplifications to
Eq.~(\ref{ADX}). First, as in Eq.~(\ref{SVnoX}), the contributions
from hadrons $X'$ produced together with the $D^*$ from the $(\bar c b)_{V-A}$
current are suppressed. Second, the contributions from non-prompt
$D^*$, that is, $D^*$ that arise from production of higher
resonances which decay into $D^*$, are suppressed. 

Before proceeding to a more detailed test, we note that
some evidence for the validity of the SV limit can be obtained 
by comparing the inclusive branching ratios measured by CLEO
\bea\label{CLEO}
Br(B\to D^{*0} X) = (0.247\pm 0.028)\,,\quad
Br(B\to D^{*\pm} X) = (0.239\pm 0.019)\,.
\eea
These numbers are summed over both charged and neutral $B$ decays,
and include semileptonic and $cd\bar u, c\bar cs$ final states.
Both semileptonic and $b\to c\bar cs$ decay mechanisms are isospin
symmetric, so they can not produce an asymmetry between the rates (\ref{CLEO}).
(Although this is manifest from (\ref{bccs}), it holds also
for the charge-averaged rates, with multiplicity factors added to account
for identical particles in the final state.)
The only source for such an asymmetry is the $b\to cd\bar u$
decay mechanism. The large $N_c$ predictions (\ref{largeN}) together with
(\ref{bccs}) imply
\bea
& &\Gamma(\bar B_d \to D^{\ast+} X^-) + \Gamma(B^{-}\to D^{*+} X^{--})\\
& & \quad
= \sum_X |F_0(X)|^2 + 3|F_1(X)|^2 + \sqrt2 \mbox{Re}(F_0(X) F_1^*(X)) + 
\sum_{i=0,1}(|h_i|^2 + |g_i|^2) + O(1/N_c)\nonumber\\
& &\Gamma(\bar B_d \to D^{*0} X^0) + \Gamma(B^-\to D^{*0} X^{-}) \\
& &\quad = \sum_X |F_0(X)|^2 + 3|F_1(X)|^2 - 
\sqrt2 \mbox{Re}(F_0(X) F_1^*(X)) + \sum_{i=0,1}(|h_i|^2 + |g_i|^2)
+ O(1/N_c) \nonumber\,.
\eea
Neglecting the $1/N_c$ corrections, the approximate equality of the 
measured rates (\ref{CLEO}) can be interpreted as evidence for a small 
ratio $F_1/F_0$, which 
coincides with the expectation from the SV limit\footnote{We neglected for
this argument the non-prompt $D^*$ production, which is itself justified 
in the SV limit.
Therefore the equality of the rates (\ref{CLEO}) provides only a consistency
check on the validity of the SV limit.}. 
Therefore we will neglect $F_1$ and compare the measured decay spectra 
against the theoretical prediction from factorization for $F_0$.

We note a subtlety in applying the combined large $N_c$ and SV limits
to inclusive decays. It is well known\cite{coleman} that in the large
$N_c$ limit widths are dominated by states with the minimum possible
number of final state mesons. Each additional meson comes at the price
of a $1/N_c$ suppression factor. Thus, naively one should include in
Eq.~(\ref{ADX}) only states in which $X$ is a single meson. However, in
the SV limit the energy available for the decay becomes arbitrarily
large and phase space effects invaliate the naive conclusion. To see
this, consider large but fixed $b$ and $c$ masses, and let $n$ be the
number of single resonances $X$ kinematicaly allowed in the decay,
that is, lighter than about $m_b-m_c$. The number of combinations of
$m$ mesons in the final state scales as $n^{am}$, where $a$ is a fixed
constant\footnote{For example, $a=1/2$ in the 't Hooft model in the
case of decays
of a heavy meson into light
mesons\cite{Grinstein:1998gc,Blok:1997hs}.}. So we see that, in
relation to the width into one light meson, the width into $m$ light
mesons roughly scales as $(n/N_c)^m$. Thus, there is no suppression of
multimeson states for $m_b-m_c$ large enough so that, roughly,
$n>N_c$.

Hence in the combined large $N_c$ and SV limits the ratio of the
inclusive and semileptonic $B\to D^*\ell\nu$ rates can be predicted in
terms of the spectral functions $v_1$ and $a_1$ for the vector and
axial currents. The spectral functions are defined through
\begin{eqnarray}
(q_\mu q_\nu - q^2g_{\mu\nu})\Pi_1^J(q^2)+q_\mu q_\nu\Pi_0^J = 
i\int d^4x\;e^{iqx}\langle0| T J^\dagger_\mu(x) J_\nu(0)\,|0\rangle\,,\\
v_1(q^2) = 2\pi\, \hbox{Im}\,\Pi_1^{J=V}(q^2)\,,\qquad
a_1(q^2) = 2\pi\, \hbox{Im}\,\Pi_1^{J=A}(q^2)\,,\nn
\end{eqnarray}
and have been measured \cite{Barate:1998uf} by the ALEPH collaboration
in $\tau$-decays up to $q^2=3.0$~GeV$^2$. Above the resonant region
the $v_1+a_1$ data displays a plateau, in excellent agreement with the
operator product expansion prediction. Therefore one can safely
extrapolate the data for $q^2>3.0$~GeV$^2$ using perturbative QCD,
$(v_1+a_1)=1.1$. This extrapolation is independent of the mechanism
responsible for factorization and therefore does not bias our analysis
in any way.

For hadronic states $X_u$ not including charmed hadrons the prediction from
factorization is 
\begin{eqnarray} \label{BDXu}
  \frac{\displaystyle \frac{\mbox{d}}{\mbox{d}q^2}\Gamma(\bar B\to D^*X_u)}
       {\displaystyle \frac{\mbox{d}}{\mbox{d}q^2}\Gamma(\bar B\to D^*\ell\bar\nu)} =
   3C_2^2|V_{ud}|^2 \Big[ v_1(q^2)+a_1(q^2) \Big] + \ldots \,,
\end{eqnarray}
where the ellipses denote terms suppressed by $1/N_c$ or two powers of
the SV expansion parameter. 
Note that although the numerical results for $B\to D^*e\nu$ in the SV limit 
are distorted by the modification in the kinematical factors, such 
modifications occur in a similar way for the inclusive and exclusive
decays and therefore cancel out in the ratio.
We emphasize that Eq.~(\ref{BDXu}) can be used to give a very
clean factorization prediction for the $\mbox{d}\Gamma(B\to D^*X_u)/\mbox{d}q^2$
decay rate with input from the $\tau$-spectral functions\footnote{The
$\tau$ decay data in this context was also used in
\cite{Aneesh} for a test of duality in nonleptonic $B$ decays.} 
and from the measured $B\to D^*\ell\nu$ form factor.

\begin{figure}
\includegraphics[width=4.in]{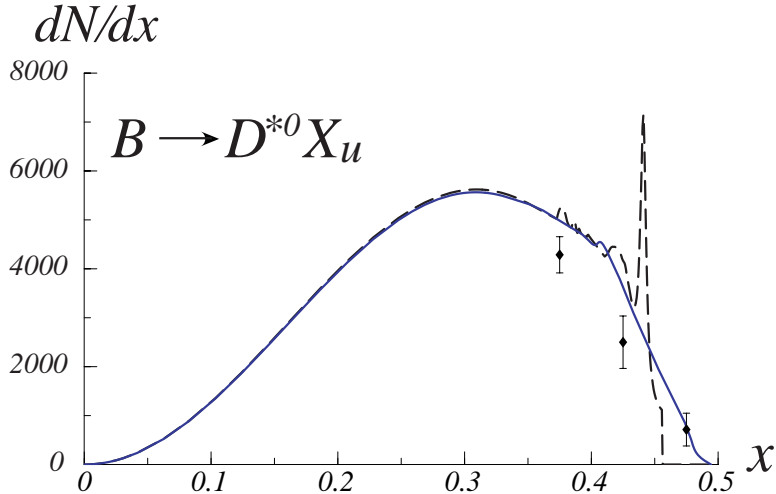}
\caption{\label{fig0} Inclusive differential decay rate 
$\mbox{d}\Gamma(B\to D^{0*}X_u)/\mbox{d}x$. The dashed line 
shows the result in
the $B$ rest frame. Here $x$ is the rescaled $D^*$ momentum,
$x=|\vec p_{D}|/(4.95\,{\rm GeV})$ and we normalize the rate as
in \cite{Gibbons:1997ag}. To compare to the
CLEO \cite{Gibbons:1997ag} data we boost to the $\Upsilon$ rest frame
which gives the solid line. Only three data points (shown) are 
available that are not contaminated by charm contributions above 
the $D_s$ threshold. }
\end{figure}

\begin{figure}
\includegraphics[width=3.in]{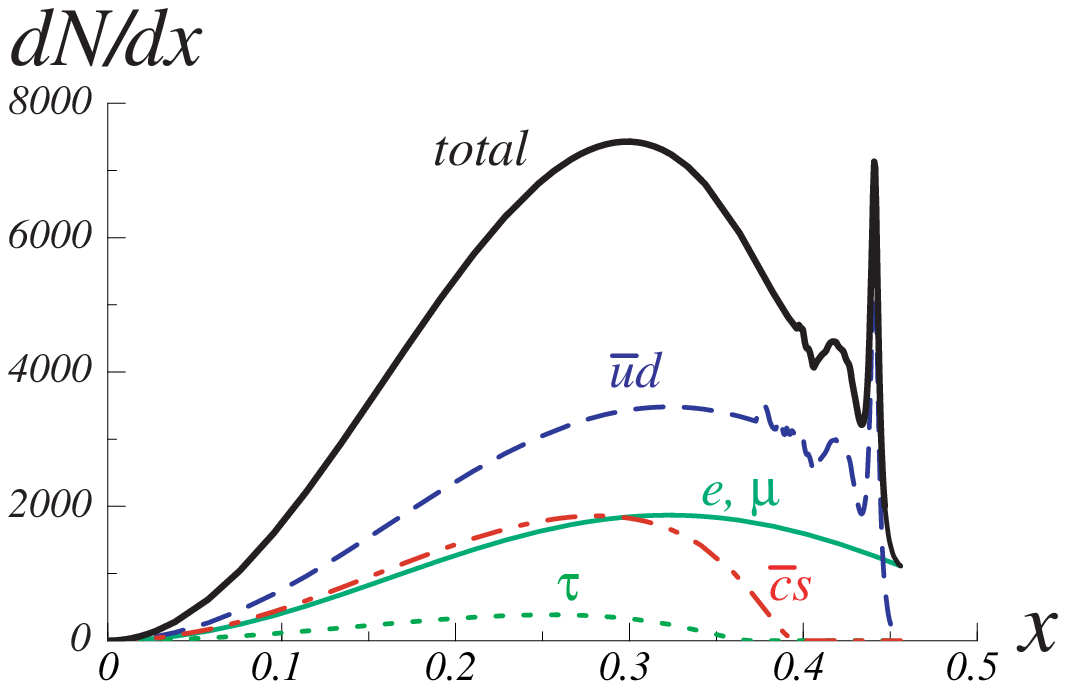}\hspace{0.5cm}
\includegraphics[width=3.in]{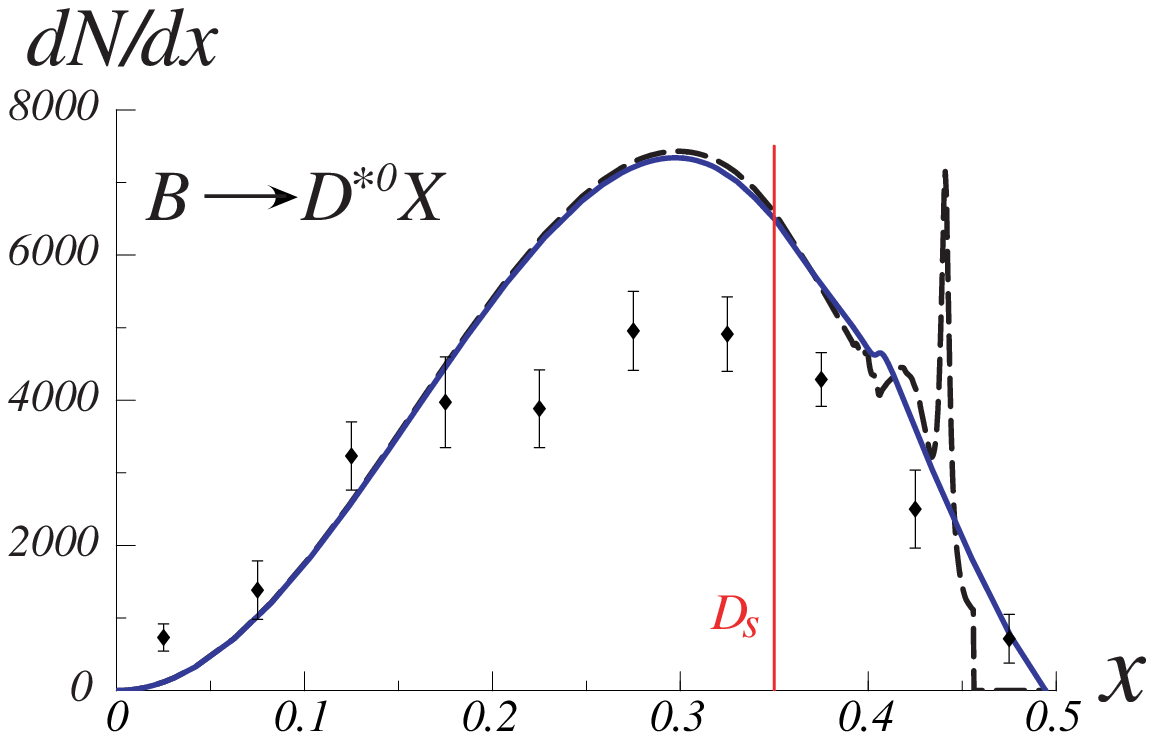}
\caption{\label{fig1} a) Breakdown of contributions to the $B\to D^*X$
spectrum in the $B$ rest frame from the $\tau$ (dotted line), $s
\bar c$ (dot-dashed line), $\{e,\mu\}$ (thin solid line), and $d \bar u $
(dashed line) final states.  The thick solid line shows the total result. b)
Inclusive differential decay rate $\mbox{d}\Gamma/\mbox{d}p$ for the process 
$B\to D^{0*}X$ using factorization from the large $N_c$ and SV limits.  The
two curves are as in Fig.~\ref{fig0}.  Charmed states in $X$ are
included using perturbative QCD to ``model'' the contributions of the
$(\bar s c)_{V-A}$ spectral density. For large $q^2$ (small $x$) this
approximation should be reasonable.}
\end{figure}

The CLEO data is presented as a function of the rescaled lab momentum of the
$D^*$ meson $x = |\vec p|/(4.95\,{\rm GeV})$. In Fig.~\ref{fig0} we show the 
boosted\footnote{In Ref.~\cite{Mannel} an analysis of the CLEO data was performed
using factorization as in Eq.~(\ref{A_Nc}) plus a model for the $X'$ contributions.
Our approach differs from this by the use of the SV limit, $\tau$-decay data,
and the inclusion of the boost to the lab frame.}
spectrum assuming that the $B$ mesons are produced
monochromatically and isotropically, with velocity $\beta=0.06$ in the lab
frame, and using
\begin{equation}
\frac{\mbox{d}\Gamma_{\rm lab}}{\mbox{d}|\vec p|}=\frac{1}{2\beta\gamma^2} \frac{|\vec p|}{E}
    \int_{s_{min}}^{s_{max}} \mbox{d}s\: \frac{1}{P(s)}\: 
\frac{\mbox{d}\Gamma_{\rm CM}}{\mbox{d}s} \: .
\end{equation}
Here $\gamma=1/\sqrt{1-\beta^2}$ and $E=\sqrt{\vec p\,^2+m_{D^*}^2}$, and
$P(s)$ and $\mbox{d}\Gamma_{\rm CM}/\mbox{d}s$ are the momentum of the $D^*$ and the decay
rate in the rest frame of the $B$ meson, respectively. The limits of
integration are $s_{\rm max}=m_B^2+m_{D^*}^2-2m_B\gamma(E-\beta |\vec p|)$
and $s_{\rm min}=\hbox{max}(0,m_B^2+m_{D^*}^2-2m_B\gamma(E+\beta |\vec p|))$.
Fig.~\ref{fig0} shows our result for the $x$ spectrum 
as a continuous curve. The CLEO data is shown
as data points. The semileptonic rate was taken from the BELLE
fit \cite{Abe:2001np} to a form factor using a unitarity constrained
parametrization \cite{Boyd:1995sq}. 
The available data is in good agreement with the factorization prediction.

Unfortunately, for large $q^2$ it is necessary to include charmed
hadrons in $X$ to enable a comparison with the experimental data that
is currently available from CLEO~\cite{Gibbons:1997ag}. If we include
charmed hadrons in the $X$ produced through a $(\bar s c)_{V-A}$
current then it becomes harder to test factorization in a clean way.
One problem is that no data for the $(\bar s c)_{V-A}$ spectral
functions is available. To make predictions for the $(\bar s c)_{V-A}$ 
current we model
the spectral functions using perturbative QCD. This neglects the
resonant structures in the region above $q^2\sim m_{D_s}^2$, but
should be fine for testing factorization at large invariant
masses. A second potential problem is that the $(\bar c s)$ and 
$(\tau\bar\nu_\tau)$
contributions require $b\to c$ form factors which are not accessible in
$B\to D^{(*)}e\bar\nu$ decay.
However, since the SV limit implies heavy quark
symmetry, this can be used to predict these form factors. Finally, an
additional source of uncertainty is introduced by the value of the
charm quark
mass. We choose $m_c=1.5\,{\rm GeV}$ in our calculations. Note that
since the $(\bar s c)_{V-A}$ current is phase space suppressed it only
contributes roughly half as much to the decay rate relative to $(\bar
d u)_{V-A}$ current. This helps to reduce the model dependence of our
$B\to D^*X$ predictions. 

In Fig.~\ref{fig1} we show a breakdown of the lepton and quark contributions 
to the total $B\to D^*X$ rate in the $B$ rest frame. 
The prediction from factorization is in moderate agreement with the data. 
The disagreement does not seem to scale with the invariant mass of the system,
however due to the large theoretical uncertainties in our calculation, 
the results in the region above the $c\bar c s$ threshold are inconclusive.
Improved measurements of the energy spectrum in Fig.~\ref{fig0}
which disentangle the states with different charm 
quark numbers would help to clarify this issue.

\section{Discussion and Conclusions} 

Data suggests that the amplitude for $\bar B^0\to D^+\pi^-$
factorizes. This can be understood via large $N_c$ counting or via large $Q$ 
as an expansion in powers of $1/Q$, the
inverse of the energy released. Thus, deviations from factorization in
this process are doubly suppressed. However, data also suggests
factorization in the exclusive processes $B \to D^* \pi^+ \pi^- \pi^-
\pi^0$ and $B \to D^* \omega \pi^-$ as well as in the inclusive $B \to
D^* X$. Neither large $N_c$ nor large $Q$ explain by themselves
these results. Indeed, large $N_c$ predicts instead a sum of
factorizable terms. On the other hand the corrections to factorization
in large $Q$ are order $m_X/Q$, and in these decays,
particularly in $B\to D^* X$, this ratio can be comparable with unity.

We have pointed out that the condition for large energy release in
$\bar B^0\to D^+\pi^-$ can be understood as a consequence of the SV limit, 
$m_b,m_c\gg m_b-m_c\gg\Lambda$. In the combined large $N_c$ and SV limits 
we can justify factorization in $B \to D^* X$ out to arbitrary invariant
hadronic mass. 

We suggest the following physical picture. The
corrections to factorization are parametrically small, of order
$1/N_c$, but depend on kinematic variables in a way that can amplify
the magnitude of the corrections if the recoiling particle is not
moving fast. The SV limit ensures that this kinematic enhancement is
absent: for cases for which we have control like $\bar B^0\to
D^+\pi^-$, the correction term is $\sim1/Q$ so it depends on the
kinematics but in such a way as to further suppress the correction to
factorization. 

This picture is additionally supported by $D$ and $K$ decay data. When
two body $D$ decay amplitudes are written in terms of weak transitions
into definite isospin states, $A_I$, and final state interaction
phases, $\exp(i\delta_I)$, it is found that the amplitudes $A_I$ do
factorize, provided one uses a modified $N_c$ counting $C_1(m_c) \sim 
C_2(m_c) \sim O(1)$ \cite{BSW}. 
The strong phases,
taken from experiment, are not small but should vanish in the large
$N_c$ limit. We interpret this as the expected kinematic enhancement
of the correction to factorization, and it suggests that factorization
fails precisely because the kinematic enhancement shows up mostly in
$\delta_I\sim 1/N_c$ rather than in $A_I\sim (1/N_c)^0$.

Is this kinematic suppression due to large velocity of the products or
large energy of the products? In the first case, which corresponds to
the color transparency argument, the suppression is expected to behave
as $1/\gamma$. In the second the suppression should be given by the larger of
$\Lambda/Q$ or $m_X/Q$, where $\Lambda$ is a typical hadronic scale,
$m_X$ is the invariant mass of state produced from the current by the
factorizing current, and $Q$ is its energy. We see that in
$K\to\pi\pi$ decays one is bound to have $\Lambda/Q\sim1$, while
$1/\gamma=2m_\pi/m_K\approx1/2$ and is parametrically suppressed in the
chiral limit. Unfortunately, the actual pion mass is too far from the
chiral limit to distinguish between the two alternatives. 

We studied in this paper the factorization predictions for $B\to D^{(*)}X$,
focusing on methods which can distinguish among the various possible
explanations for factorization. In Sec.~II we derived large $N_c$
relations among isospin amplitudes, which lead to observable predictions
among decay rates. 
Imposing additionally the $SV$ limit gives even more relations, since certain 
amplitudes allowed in the pure $N_c$ limit are suppressed. 
The large $Q$ limit by itself gives predictions similar to the combined limit 
of large 
$N_c$ and $SV$, however these predictions should fail outside of a limited 
kinematical range. We presented several such predictions that can be tested
experimentally. Using available data from BELLE and Babar we discussed such
predictions for the modes $B\to D^{(*)}\pi\pi$, $B\to D^{(*)}K\bar K$ and
$B\to D^{(*)}\bar D^{(*)} K$.

In Sec.~III we calculated the inclusive decay rate 
$B \to D^* X$ in the combined limit of large $N_c$ and $SV$. In this limit,
factorization is expected to work out to arbitrary invariant hadronic mass, 
in contrast 
to the predictions from large $Q$ factorization in which factorization 
breaking corrections should scale as $m_X/Q$. The cleanest theoretical 
prediction is for the $B\to D^*X_u$ decay, for which 
we can use $\tau$-decay data to extract the spectral function in the resonance 
region\footnote{We take this opportunity to comment on the recent measurement
by BELLE \cite{BelleICHEP}
of the ratio $Br(B^-\to D_2^{*0}\pi^-) / BR(B^-\to D_1^0\pi^-) = 0.89 \pm 0.14$, 
which has been interpreted as a test of factorization. However, as pointed out
in Ref.~\cite{Leibovitch} this ratio depends sensitively on unknown subleading
Isgur-Wise functions, which can accomodate values in the range $0-1.5$ within
factorization.}. The comparison of the prediction with the CLEO data is shown in 
Fig.~\ref{fig0} and shows good agreement. Inclusive data for this process 
is available over a much larger kinematic range, however the data then 
includes contributions from the $(\bar s c)_{V-A}$ part of the current. 
No data is available for the spectral function of this current, and we 
therefore have to rely on perturbative calculations. The result is shown 
in Fig.~\ref{fig1}, however the large theoretical uncertainties preclude us 
from drawing definite conclusions. 

We emphasize that it would be desirable to separate the $b\to cd\bar u$ 
and $b\to c\bar cs$ final states in the $B\to D^{(*)}X$ data, including separate 
measurements for $B$ and $D$ charge eigenstates. The methods described in this paper
would then allow a more detailed study of factorization in this decay, and should help
shed light on the mechanism underlying the observed factorization 
in $B$ decays.

\acknowledgments{
We are grateful to Vivek Sharma for many useful discussions.
This work is supported in part by the Department of Energy
under contracts No.\ DOE-FG03-97ER40546 and DE-FG03-00-ER-41132.
}

\end{document}